\documentclass[conference]{IEEEtran}
\usepackage{cite}
\usepackage{amsmath,amssymb,amsfonts}
\usepackage{algorithmic}
\usepackage{graphicx}
\usepackage{textcomp}
\usepackage{xcolor}
\usepackage{hyperref}
\def\BibTeX{{\rm B\kern-.05em{\sc i\kern-.025em b}\kern-.08em
    T\kern-.1667em\lower.7ex\hbox{E}\kern-.125emX}}
\begin{document}

\title{Designing an Intelligent Parcel Management System using IoT \& Machine Learning \\
\thanks{}
}
\author{\IEEEauthorblockN{Mohit Gupta\IEEEauthorrefmark{1},
Nitesh Garg\IEEEauthorrefmark{2}, Jai Garg\IEEEauthorrefmark{3},
Vansh Gupta\IEEEauthorrefmark{4}, and Devraj Gautam\IEEEauthorrefmark{5}}
\IEEEauthorblockA{Electronics and Communication Engineering Department,\\
Dr. Akhilesh Das Gupta Institute of Technology \& Management, \\
New Delhi, India\\
\IEEEauthorrefmark{1}guptamohit1504@gmail.com,
\IEEEauthorrefmark{2}nitesh.garg049@gmail.com,
\IEEEauthorrefmark{3}jaigarg2@gmail.com,
\IEEEauthorrefmark{4}vgvansh25@gmail.com,\\
\IEEEauthorrefmark{5}devrajgautam10@gmail.com}}

\maketitle

\begin{abstract}
Parcels delivery is a critical activity in railways. More importantly, each parcel must be thoroughly checked and sorted according to its destination address. We require an efficient and robust IoT system capable of doing all of these tasks with great precision and minimal human interaction. This paper discusses, We created a fully-fledged solution using IoT and machine learning to assist trains in performing this operation efficiently. In this study, we covered the product, which consists mostly of two phases. Scanning is the first step, followed by sorting. During the scanning process, the parcel will be passed through three scanners that will look for explosives, drugs, and any dangerous materials in the parcel and will trash it if any of the tests fail. When the scanning step is over, the parcel moves on to the sorting phase, where we use Qr codes to retrieve the details of the parcels and sort them properly. The simulation of the system is done using the blender software. Our research shows that our procedure significantly improves accuracy as well as the assessment of cutting-edge technology and existing techniques.
\end{abstract}


\section{\textbf{Introduction}}
Today’s as the population increases, more and more people uses transports like railways, and flights. They carry their luggage and for the safety purposes all the peoples and their luggage needs to be passed through a safety check which checks for any dangerous or explosive material in the luggage. There are scanners, but still, many police officers or security professionals are assigned to manually checks for each of the luggage to check for any dangerous or explosive materials in the luggage. People have to wait there for checking in, which becomes the one of the main reason of crowds at these stations.

The Indian railway network is the world's third biggest. The railway industry is known as one of the largest railway networks under single control, thanks to the government's concentrated attention and investments in strengthening infrastructure. Aside from transporting over 10,000 people every day, over 7,000 freight trains transport over three million metric tonnes of freight per day. The average freight charge per metric tonne across the country in fiscal year 2020 was 1.58 Indian rupees per kilometre. Revenue throughout the sector has steadily increased throughout the years. In fiscal year 2018, freight rail transport income was expected to reach close to 2.5 billion US dollars across the country. With increased engagement from both government and non - governmental enterprises, both local and international, freight traffic is expected to develop fast, potentially leading to greater job possibilities and income production from the industry [1].

As the parcel delivery in Indian railway is very less costly, more the amount of people uses railways for delivering their parcels from one state to another. From the recent statistics around 65\% of the revenue of Indian railways is from freight transportation. With the advantages there are still some loopholes and disadvantages here. It is seen that many accidents happens day by day in Indian Railways, there may be numerous reasons for it, like proper checking is done, people crosses the danger line at railway platforms, walking on the track and many more. One of the main reason of accident in Indian railways is Terrorist incidents. Luggage transportation with explosives. As there are still some things missed out in regular checking due to manual checking. Transportation of drugs, explosives, dangerous things are illegal as it can be the reason of some misshaping.
The Global Terrorism Index assesses the consequences of terrorism, including the number of lives lost, injuries, property destruction, and psychological consequences. It is a weighted combination that grades nations according to the impact of terrorism from 0 (no influence) to 10 (great impact). The terrorism index in India is approximately 7 [2]. We can see the Terrorism Index variation in India from 2010-2020 in the Fig. 1. In some cases, the explosives like things are transported through Indian railways as parcels. 

\begin{figure}[htp]
    \centering
    \includegraphics[width=8.5cm, height=4cm]{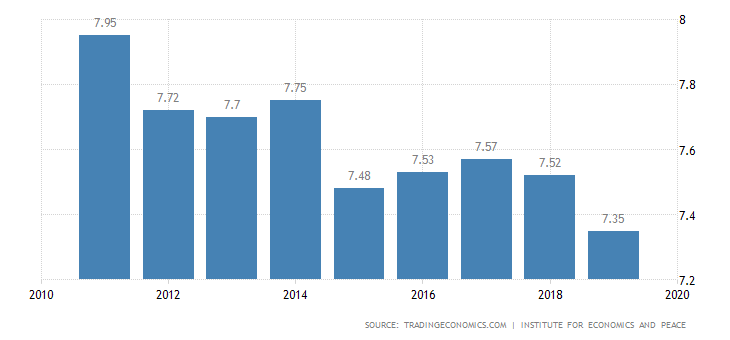}
    \caption{The terrorism index in India from 2010-2020}
    \label{fig:The terrorism index in India}
\end{figure}

So to prevent this Indian Engineers are doing research in this field, about how we build a solution which will minimize the human intervention and make a system which will efficiently scan and sort the parcels.

As advancement in Machine learning and Artificial Intelligence, numerous technology and algorithms are built in foreign for this purpose. Sometimes, many parcels are lost in the way, or does not reach to the destination. For this, an efficient real-time parcel tracking and monitoring also needed.  
In this paper, we have discussed about the solution that can help the railway authorities of India, for better and efficient way for handling daily parcels scanning and sorting process with minimal cost and manual intervention. The solution is complete IoT based solution with the implementation of machine learning. We have also build the web application for getting the real-time parcel details. Our solution consist of two phases, first is scanning and the second one is sorting. In the scanning phase we’d used RF-ID for record entry of the parcel, metal detectors for any metallic object filtration, and IR thermal detectors to detect explosives inside the parcel and we plot the heat map on the CT-scanned image to show explosives. 
In the sorting phase, we’d used QR codes for sorting purpose. We store the details like Dimensions, weight \& Address of the parcel in the QR-code. Then we’ll scan the QR code and get the details of the parcel and sort them accordingly. This can reduce the labor time for sorting the parcel in their respective areas. The more details about the proposal will be explained in the section III.

\section{\textbf{LITERATURE REVIEW}}
Tan, Z. et al. [3] examines the parcel sorting optimization problem in an online store center where goods are waiting to just be categorized and handed over the packages, picking stations, and AGVs are assigned along with the goal of decreasing the time it takes to process the final item, so that a comprehensive warehouse sorting operations plan may be produced. Based on the problem, a mixed-integer linear programming model is created, and the particle swarm optimization (PSO) algorithm is used to solve it. The sorting conveyor and AGVs using a pooling strategy are used to handle sorted packages in the vertical sorting system. Introduced A mixed-integer linear programming model is developed to reduce overall sorting time and to assign destinations, pickup stations, and AGVs.

R. T. Yunardi, Winarno and Pujiyanto [4], In their article, they constructed a computer vision system capable of determining the volume of parcel boxes. We need to know the length, breadth, and height to acquire this value. They employed two webcam cameras to determine the dimensional scale of the parcel by computing the pixels collected on camera and comparing them for calibration. The 2D picture is made up of two camera images, one vertical and one horizontal. After obtaining the width, length and height of the parcels box, a multiplication software will be utilized to determine the volume result. It will be automatically separated on the conveyor belt. The device can recognise boxes with a length and width of 1 - 15 cm and a height of 5 - 20 cm. The average accuracy achieved in giving information about the levels of the parcel size is 87.5\%. 

Tan et al. [5], Their research optimizes the sorting scheme for a specific type of the double-layer sorting machinery that employs a "component-sorting" technique. The process of sorting many parcels transported to the same location using the same tray is referred to as "component-sorting." An integer programming paradigm is used to solve the problem. Then, to solve the problem quickly, we offer an efficient variable neighbour taboo search. This work develops a type of algorithm that combines taboo searches with multi objective search to provide a better sortation scheduling method in less time to help the equipment operator make judgments. In comparison to CPLEX, the technique takes less time to find a solution and can handle huge cases quickly. As a result, VNTS2 is more suited to addressing this issue.

L. Zi and B. Gao in [6] a newly-designed model is formulated to optimize both system throughput and cost. Furthermore, we have built an M/M/s queuing model to estimate the throughput of the system. Besides, we have designed a global search algorithm to solve the models. The impact of various system components on system throughput, such as the length-width ratio, the quantity and arrangement of terminals and receptacles, and the number of autonomous guided vehicles, are studied using numerical experiments.

Khir, R et al. [7] presented strong planning models that send parcels to sort equipment while safeguarding parcel carriers from various sources of demand unpredictability. We establish the computational feasibility of the suggested models using realistic-sized examples based on industrial data, as well as their utility in giving sort plan options that trade off operating costs and levels of resilience. The results demonstrate that such restricted flexibility might assist lower the cost of resilience, particularly in Day sorts sessions, which generally process tiny parcel quantities bound for a variety of proximate and local destinations.

It was recommended in [8] by Yonghoon Choi et al. that a UHF RFID system be used to track and manage parcels from the time they are picked up at the pick-up post office until they are delivered at their destination post office. At the Uijeongbu mail sorting facility, their system was tested from the pick-up post office to the delivery, covering the whole process of package handling. Each station in the mail sorting facility has a multi-tag identification RFID technology that can scan and sort the parcels automatically without the need for typing.

Zainudin, Juanita et al. [9] proposed that the result of their invention, delivered packages may be tracked by computer using a QRcode or tracking number, which will alert receivers through SMS. The package was managed and tracked via a web-based service. This system was developed using PHP and JavaScript in Adobe Dreamweaver, using MySQL as the database backend. Their Parcel Tracking System increases efficiency in terms of information storage and notification messages. They found that users require electronic systems that allow them to manage their possessions. Parcel Tracking System is also used at KUPTM to reduce human mistakes and give administrative staffs a new way to locate and save data. There's also evidence of a system's speedy transmission capabilities to the receiver. It is possible to make a number of changes to the proposed system in order to boost its capabilities.

Using the notion of service quality and customer happiness in the adoption of relevant technologies, Tang et al. [10] conducted their research in this study. According to their paper, the service quality of smart parcel lockers can be broken down into five categories: service pricing; service dependability; convenience; fault management abilities; and service diversity. User satisfaction with smart package locker services was examined using 272 genuine Chinese questionnaires. To check the reliability and validity of the statistical analysis, a confirmatory factor analysis was done (CFA). Customers' satisfaction levels were also correlated, as was a regression analysis. Service pricing has no beneficial impact on consumer satisfaction, but the other four elements have a favorable impact. Since the study is focused on parcel lockers, it will have a substantial influence on the future development of last-mile services.

Ahmad, M et al. [11] study report compares the efficiency of RFID technology in the courier services business to the present QRcode method. By analyzing mistake rates and the time it takes to scan RFID tags, they can determine the impact of RF-ID on inventory accuracy in the courier warehouse and distribution center system. Using RF-ID, they look at how inventory management systems may be improved by lowering the amount of time spent and increasing the number of products that are accurate. Documentation and research papers have been used to conduct this study. RF-ID and bar codes are two technologies that have been extensively studied. According to the study, RF-ID can scan more quickly and improve inventory accuracy more effectively than bar code scanning systems.

H, Medhat \& Awadalla, A [12], this study provides a feasible method for online cargo tracking. In order to figure out exactly where it is at any one time, from when it was sent to when it was delivered, it uses a tracking system. An online database-driven application has been built for the system, which facilitates its maintenance and gives relevant information about the cargo. In addition to the database, the created system also includes a user-friendly interface for tracking shipments and knowing whether they have been damaged or lost, as well as their whereabouts until they reach their final destination. Verification of the entire system have been performed, and the findings have been encouraging.

However, it does not confine itself to a single study field or explosive type. During this review [13], we'll be looking at what's changed in the last five years, and referring the reader to previous reviews where necessary. As can be seen, virtually all of the analytical methods examined in this study have undergone major adjustments in the previous five years to enhance one or more elements of their working procedures. On account of the inherent problems connected with explosive identification, such as the low volatility of explosive vapours and concealment concerns as well as the harm caused by a false positive reaction, more development will be required. We've made huge strides in the mobility of equipment; formerly fixed-position spectroscopic methods like IMS and mass spectrometry are now field deploy-able. One of the most controversial aspects of Terahertz imaging is that it may be used to detect bombs and other hidden items under clothing and in baggage. During the recent decade, nanotechnology has grown in importance, with researchers looking for ways to incorporate new nano-based elements into established technologies in order to increase their sensitivity, selectivity, and mobility. 

Hernández-Adame et al. [14]. Using the Monte Carlo technique, a neutron source 241AmBe was used to build an explosive detection system. Moderators in the design included light water (as well as paraffin, polyethylene, and graphite). There were two types of detectors used: NaI (Tl) and HPGe. For the optimum performance, light water as a moderator should be combined with HPGe as a detector. The MCNPX code was used to develop an explosive detection system. When RDX and urea were employed as neutron capture samples, four materials were used in the moderator, and two detectors were used to detect the photon generated by this process. Main conclusions based on evidence: 
\begin{itemize}
  \item \textbf{(n, \(\gamma\))} reactions are most common near the neutron source, which is positioned about half a centimeter below the cell's sample surface.
  \item Because of the neutron capture in O and the photon capture in C, a greater quantity of \textbf{\(\gamma\)-rays} are produced when light water is employed as a moderator, which is in agreement with the elemental concentration in explosives.
  \item In explosives, urea can be employed as a line-base between explosives and organic materials since the characteristic \textbf{\(\gamma\)-rays} generated during neutron capture are greater than the photons produced in urea.
\end{itemize}

\section{\textbf{PROPOSED SYSTEM}}
After reviewing all the research published on this regard, we have built a complete solution which will cover almost all of the loopholes of the previews research. Our main aim is to save the time, detect impermissible objects inside the parcel and also to reduce human efforts. We have used RF-ID (radio frequency identification) together with QRcode so as to reduce the chances of missing a parcel when being scanned as the QRcode needs to be facing the camera for being read and RFID can be read by a RFID reader irrespective of the orientation of the parcel moreover RFID is much secure than the QRcode and even holds more data (QRcode scanner even holds the weight sensor beneath the conveyer and sharp sensor for calculating the size of the parcel as well). 

\begin{figure}[htp]
    \centering
    \includegraphics[width=7cm, height=4cm]{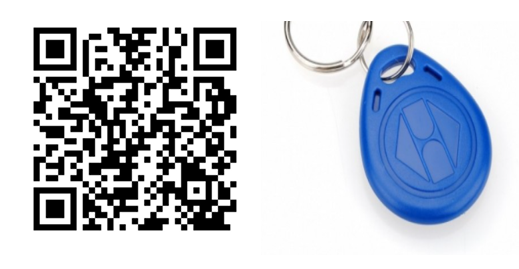}
    \caption{QRcode \& RFID Tag}
    \label{fig:QRcode}
\end{figure}

Our system consist of majorly two phases:
\begin{itemize}
  \item Scanning Phase.
  \item Sorting Phase.
\end{itemize}

\subsection{\textbf{Scanning Phase}}\label{AA1}
We have used three different types of scanners and each one of them have their distinct role. They all are present together in a scanning system where they get passed if found permissible, otherwise if any scanner out of four detects any impermissible object inside the parcel then the parcel will move towards a separate drop location by help of an another conveyor belt and will be dumped to the dumping zone for physical handling.

\subsubsection{\textbf{Metal Detection}}\label{AA2}
The first scanner scans for the metal detection. We have built a low cost metal detector as shown in Fig. 3 which is connected through buzzer and raspberry pi for processing the input and gives the output on the buzzer. Metal detector is most sensitive to things in which a current can flow in the plane of the coil, with the response proportional to the area of the current loop in that item. As a result, a metal disc in the plane of the coil will produce a significantly stronger response than a metal disc perpendicular to the coil [15].

\begin{figure}[htp]
    \centering
    \includegraphics[width=7cm, height=4cm]{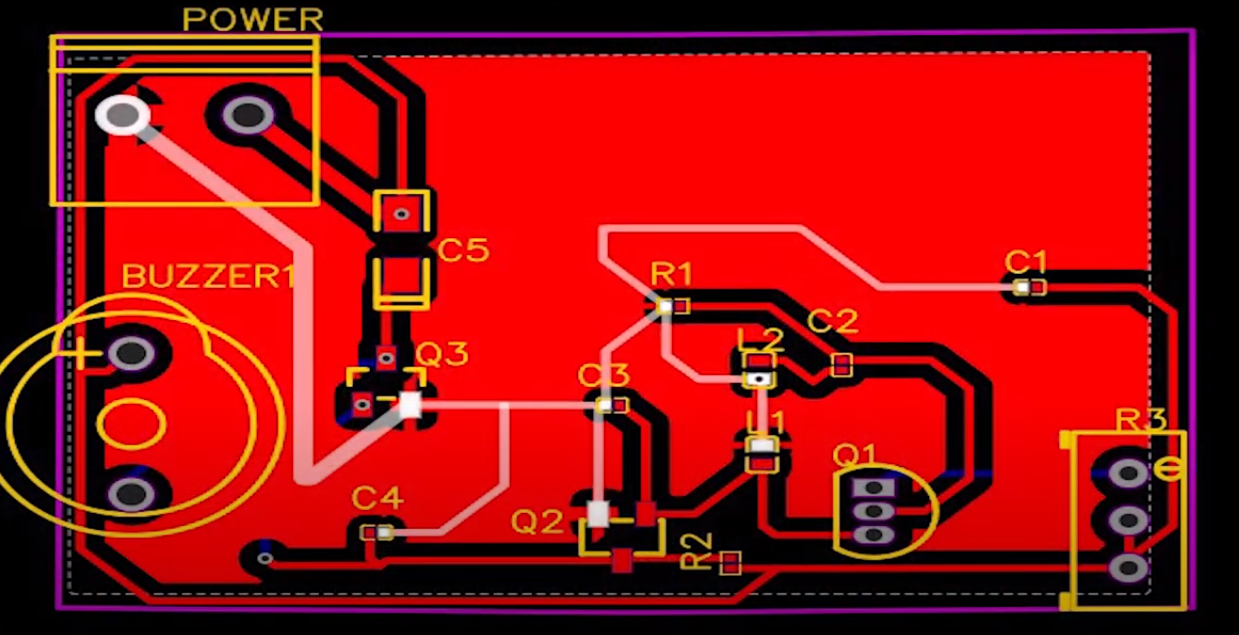}
    \caption{PCB layout of Metal Detector}
    \label{fig:pcb layout of metal detector}
\end{figure}

When electricity flows through a coil, it creates a magnetic field. A changing magnetic field, according to Faraday's law of induction, will result in an electric field that opposes the change in magnetic field. As a result, a voltage across the coil develops that opposes the increase in current. This is known as self-inductance. The slight relative variations in coil inductance caused by the presence of surrounding metals are used for metal identification. When a metal is detected by the metal detector in the parcel, then the buzzer beeps, and the parcel is discarded for manual checking, else it will pass through the second scanner, which will detect any objectionable things like gun, knife, or any sharp object in the parcel. This is done using the X-ray.

\subsubsection{\textbf{X-Ray Detection}}\label{AA3}
The absorption of X-rays differs based on the material. The rates of absorption are influenced by the material, thickness, and concentration of the substance. When a conveyor belt's internal wire ropes are strained or damaged, the highly permeable X-ray strength of the defect location is greater than the standard portion due to the defect position's lower absorption, as well as the X-ray intensity obtained by sensitive opto electronic components is high [16]. The X-ray images received from the X-rays detector, is passed through a machine learning model, which process the image and detect if it contains any of the objectionable images. The X-ray image example is shown in Fig. 4. The machine learning model is trained on around 1GB of x-ray images data-set. If the parcel contains any objectionable object, then it will redirected to the dump section.

\begin{figure}[htp]
    \centering
    \includegraphics[width=7cm, height=10cm]{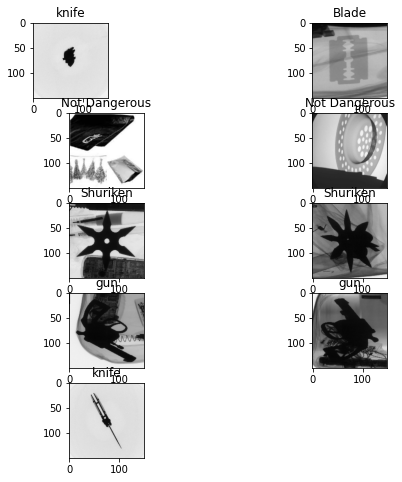}
    \caption{X-RAY detection image from ML Model}
    \label{fig:X-RAY detection image from ML Model}
\end{figure}

\subsubsection{\textbf{Explosive Detection}}\label{AA4}
For this stage, the IR thermal detection is used. For explosives and drugs detection we are using IR thermal detection. Infrared thermography is a contact-less method of capturing and interpreting thermal information from a surface or object [17]. The scanning of temperature is made feasible by the fact that a grey body generates electromagnetic radiation in the infrared spectrum, with a wavelength ranging from 0.7m to 1mm [18]. It is feasible to measure the surface temperature of an object by utilizing its property that IR radiation greatly depends on the temperature of the emitting body [19]. To accomplish thermal imaging based on emitted radiation, an IR camera goes through a series of steps that include optical focusing of radiation, filtering of the IR spectral band, detection of the IR ray, conversion of the energy into electrical voltage, and signal processing [20].

\begin{figure}[htp]
    \centering
    \includegraphics[width=8.5cm, height=2cm]{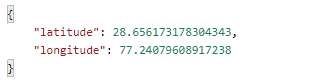}
    \caption{GPRS Coordinates format from API response}
    \label{fig:img1}
\end{figure}

The machine learning model then creates a heat map from the IR images, and find out which parcel contains drugs or explosives and discards the package. 

\subsection{\textbf{Sorting Phase}}\label{AA5}
Sorting of parcel will be done on three basis which is purely user choice which is something new. Sorting will be done by following basis:- 
\begin{enumerate}
  \item Weight 
  \item Dimensions
  \item Zone
\end{enumerate}

Reaching the sorting zone, parcels will be sorted based on nature like metallic or non metallic, delivery location, size, weight, fragility. Further there will be a sharp sensor, each aligned with a drum. On detecting a parcel via the sharp sensor, the drum will push the parcel to the specific container based on either nature.

\begin{figure}[htp]
    \centering
    \includegraphics[width=7cm, height=3.5cm]{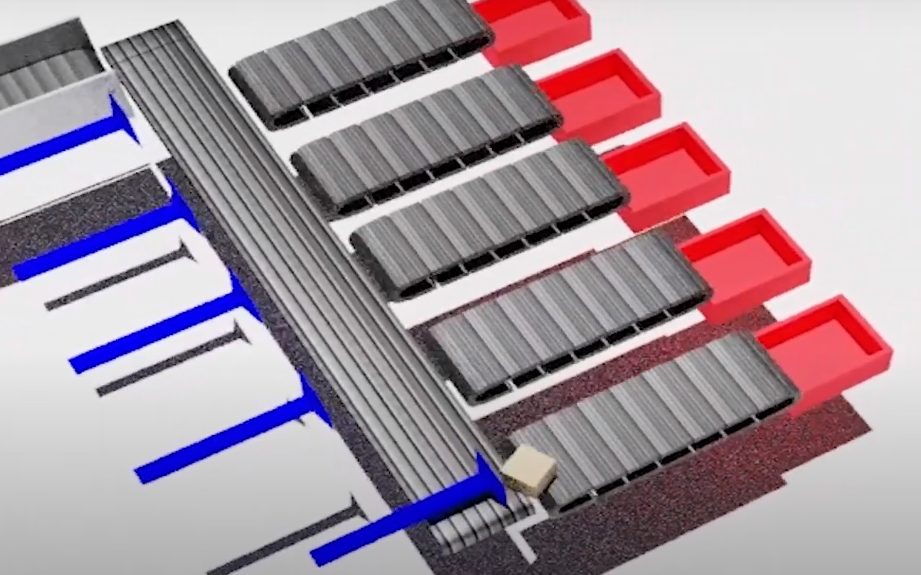}
    \caption{Sorting Simulation of parcel}
    \label{fig:Sorting of parcel}
\end{figure}

This system also have a real time tracking system in the web interface with the help of which users can track their parcels in real time. The coordinates of the parcel will be updated on the firebase realtime database periodically. Let say, parcel have to reach delhi from rajasthan, then checkpoints will be set between the journey, so when the parcel reaches the checkpoint, the coordinates of the checkpoint will be updates on the firebase and then utilized by the interface to show on the map.

\begin{figure}[htp]
    \centering
    \includegraphics[width=7cm, height=4cm]{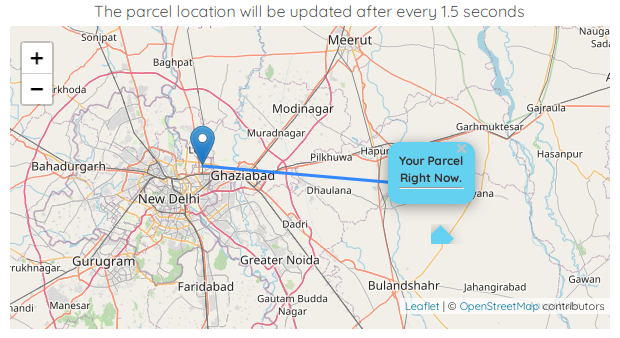}
    \caption{Live Tracking of Parcel}
    \label{fig:tracking}
\end{figure}

\section{\textbf{Implementation}}
\subsection{\textbf{Data Collection}}
OpenCV is a Py computer vision toolkit designed to make programs computational. This library contains a programme that makes use of the computer's webcam. Once imported, a reference to the computer's standard webcam was formed. It created an item for VideoCapture. The object's argument is the room index for our index 0. This adjustment allows us to get the required results by presenting considerably more information on the computer. Data may contain a range of information, like the range between the framed item and the camera at a certain point. The motion of the cameras in an image is transformed conversion process in black and white hues is another characteristic of computer vision. The processor can identify pictures in numeric arrays and the noise portion of any number in that array. This is the finest computer vision has to offer.

The model function, which also creates a matrix, is now ready to fit the data vector containing the binary values that correspond to each data set's labels. The 'concatenate' functionality in the Num Py library comes first, followed by the 'to categorical' feature in the 'keras.utils' package.

\subsection{\textbf{Structure of the CNN}}
CNNs are made up of 2 convolutional layers, which serves as the neural network's fundamental unit. A conversion layer in the matrix executes a sequence of linear detection and monitoring. The feature map refers to the image processing methods that result from the concatenation functions image processing techniques that come from the input picture or image from the previous layer.

\begin{enumerate}
  \item \textbf{Two levels pooling} – In the feature map, replace the area value with the data for the region. The grouping function MaxPooling used within the system. This replaces a precisely defined zone with the greatest area, resulting in a more compact layout while retaining the most important information. \textbf{Four drop-out layers} - The drop-out layer is particularly important in neural networks for preventing overpackaging. This layer assumes that some level weights will be disregarded and many networks containing features will be simulated with fewer.
  \item \textbf{A single layer of flattening} — It must be feasible to use entirely connected layers after specific convolutional layers. As a result, the tensor formed from the convolution layers becomes a 2D vector.
  \item \textbf{The dense layer: An NLP network's standard layer}, with weight nodes and compensations distributed across three dense (compact) layers (bias). The grouping layers (the size of the frame for each level) and knock levels (the ignition rate for every layer) are usually put after the convolution layers. The flattened layer is then shown. Then we have layers that are completely connected (each layer's number of neurons is specified). The neural network's design and number of layers are described in a table.
\end{enumerate}

\section{\textbf{Results}}
Our solution is a low-cost, high-efficiency system that solves the problem while reducing manual intervention in package scanning and sorting. Airports and railways may successfully use this to speed up their parcel management systems.

\begin{table}
\begin{center}
\caption{Classification Report}
\resizebox{8cm}{1.2cm}{$
\begin{tabular}{ | l | r | r | r | r | } 
\hline
  & PRECISION & RECALL & F1-SCORE & SUPPORT \\
\hline
Blade & 0.87 & 0.87 & 0.87 & 15\\
\hline
gun & 0.78 & 0.82 & 0.80 & 17\\
\hline
knife & 0.82 & 0.93 & 0.87 & 15\\
\hline
Shuriken & 0.96 & 0.91 & 0.94 & 57\\
\hline
Non-Dangerous & 0.87 & 0.87 & 0.87 & 15\\
\hline
\end{tabular}$}
\end{center}
\end{table}

Blade, gun, knife, shuriken, non-dangerous were chosen as five categories of things to observe the program's performance. Following are the results of the testing of the large datasets that make up these classes, as shown in Table I. According to the classification report, the precision and recall for each object are rather excellent, indicating that our model has a very low false positive rate and good accuracy.

\begin{table}
\begin{center}
\caption{Model Accuracy}
\resizebox{7cm}{0.8cm}{$
\begin{tabular}{ | l | r | r | r | r | } 
\hline
  & PRECISION & RECALL & F1-SCORE & SUPPORT \\
\hline
accuracy & & & 0.89 & 119 \\
\hline
macro avg & 0.86 & 0.88 & 0.87 & 119 \\
\hline
weighted avg & 0.89 & 0.89 & 0.89 & 119 \\
\hline
\end{tabular}$}
\end{center}
\end{table}

\section{\textbf{Conclusion}}
Parcel Management is very important and crucial tasks which is managed by railways and airport security for the safety and delivery of parcels safely to their destinations. We suggest a novel way to make this process smooth, cost efficient and less human intervention system that will help railways and airports to manage parcels from the entry in the database to delivery to the destination. This proposal is a fine blend of IoT and Artificial Intelligence in effectively solving the real-time problem. We suggest making use of this system with some standard improvements will reduce time and man power in scanning and sorting of parcels. We further suggest using the web interface to gain a real-time parcel tracking and managing the sensors, will greatly increase the performance, feedback and profit of the organization. As demonstrated in TABLE II, our model has greater accuracy and precision than the stated previously model.

\end{document}